\input{aipcheck}

\documentclass[
    ,final            
  ]
  {aipproc}

\layoutstyle{8x11single}

\begin{document}

\title{Charged charmonium-like structures and the initial single chiral particle emission mechanism\footnote{Presented by T. Matsuki at XIth Quark Confinement and the Hadron Spectrum held at St. Petersburg, Russia on Sept. 8.}}

\classification{14.40.Rt, 13.25.-k, 13.25.Gv, 12.38.Bx}
\keywords      {exotic mesons, Hadronic loop, two-chiral particles emission}

\author{Dian-Yong Chen}{
  address={Nuclear Theory Group, Institute of Modern Physics, Chinese Academy of Sciences, Lanzhou 730000, China}
}

\author{Xiang Liu}{
  address={School of Physical Science and Technology, Lanzhou University, Lanzhou 730000, China}
}

\author{Takayuki Matsuki}{
  address={Tokyo Kasei University, 1-18-1 Kaga, Itabashi, Tokyo 173-8602, Japan}
}

\begin{abstract}
This paper summarizes what we have done so far to explain charged charmonium/bottomonium-like structures using a hadronic triangle diagram which we call the initial single chiral particle emission mechanism. We discuss processes like $A\to A'+P+P'$ where two chiral particles $P$ and $P'$ are emitted from $A$. In the intermediate we consider a hadronic one-loop diagram in which $D^{(*)}/D_s^{(*)}/B^{(*)}/B_s^{(*)}$ are included to explain some enhancements experimentally observed. Using this mechanism we explain some of the exotic enhancements and predict a couple of enhancement structures.
\end{abstract}

\maketitle


\section{Introduction}

There have been abundant observations of charmonium-like states $XYZ$ in the past decade, which have been reported by BaBar, Belle, CLEO-c, CDF, D0, CMS, LHCb and BESIII. Experimental observations of these experimental groups have stimulated theorists' extensive interest (see recent reviews \cite{Liu:2013waa} and \cite{Esposito:2014rxa} for more details).

In this talk, we introduce an important mechanism existing in the hidden-bottom/charm decays of higher bottomonia/charmonia, which is named the initial single chiral particle emission (ISChE) mechanism. Utilizing this mechanism, we have given plentiful novel phenomena relevant to charmonium-like and bottomonium-like structures, which can be accessible by experiments.

There are a couple of ways to explain a peak structure as a kinematical origin other than ours among which threshold cusp effects have a long history. For instance, Wigner considered a process $A+B\to A'+B'$ and tried to explain a peak structure near the threshold \cite{Wigner:1948zz}. That is, the energy is close to a sum of two intermediate masses, then the amplitude behaves like $1/\sqrt{|1-(m_1+m_2)^2/s|}$, and there occurs a singularity at $s=(m_1+m_2)^2$. He used a quantum mechanical argument which can be applied to a field theoretical problem, too.
Cabibbo revived this idea and applied it to explain a cusp of the process $K^+\to\pi^+\pi^0\pi^0$ \cite{Cabibbo:2004gq} using a bubble-type one loop diagram with two vertices.
Recently Bugg initiated a way to describe a threshold cusp effect using the dispersion relation \cite{Bugg:2008wu} and have tried to explain $Z_b(10610/10650)$ \cite{Bugg:2011jr}. As an alternative approach to explain a fictitious peak structure, we propose another method, the ISChE mechanism in which we consider the process $H_1\to H_2+P+P'$ where two heavy particles, mostly charmonia/bottomonia, $H_1$ and $H_2$ and two chiral particles $P$ and $P'$ are introduced.

\section{Lessons from hidden-bottom dipion decays of $\Upsilon(5S)$}

In 2011, the Belle Collaboration reported two charged bottomonium-like structures $Z_b(10610)$ and $Z_b(10650)$
in the $\Upsilon(5S)\to h_b(mP)\pi^+\pi^-$ ($m=1,2$) and $\Upsilon(5S)\to \Upsilon(nS)\pi^+\pi^-$ ($n=1,2,3$) decay processes \cite{Collaboration:2011gja}. The typical properties of these two $Z_b$ states are that $Z_b(10610)$ and $Z_b(10650)$ have charged structures and are near the $B\bar{B}^*$ and $B^*\bar{B}^*$ thresholds, respectively. 
Before this observation, Belle once studied the hidden-bottom dipion decays of $\Upsilon(5S)$ and indicated that 
the branching ratios of $\Upsilon(5S)\to \Upsilon(nS)\pi^+\pi^-$ ($n=1,2,3$) are larger than the dipion transition rates between the lower members of the $\Upsilon$ family by two orders of magnitude \cite{Abe:2007tk}, which is a puzzle of $\Upsilon(5S)$ decays. For solving this puzzle, the rescattering mechanism \cite{Meng:2007tk} and tetraquark state $Y_b$ near $\Upsilon(5S)$ \cite{Ali:2009pi} were proposed. Although these two mechanisms can explain why the $\Upsilon(5S)\to \Upsilon(nS)\pi^+\pi^-$ decays have larger decay rates, they failed to describe the experimental data of the $\Upsilon(2S)\pi$ and $\pi\pi$ invariant mass spectra of $\Upsilon(5S)\to \Upsilon(2S)\pi^+\pi^-$ \cite{Ali:2009es,Chen:2011qx}. Thus, the authors of Ref. \cite{Chen:2011qx} indicaed that there exists a new puzzle in $\Upsilon(5S)\to \Upsilon(2S)\pi^+\pi^-$ decay. 

Later, Chen, Liu and Zhu realized the relation of two experiments mentioned above, which is crucial to solving this new puzzle in $\Upsilon(5S)\to \Upsilon(2S)\pi^+\pi^-$ decay \cite{Chen:2011zv}. By introducing the contribution from two $Z_b$ structures in $\Upsilon(5S)\to \Upsilon(2S)\pi^+\pi^-$ and considering the interference effects of different mechanisms, the $\Upsilon(2S)\pi$ and $\pi\pi$ invariant mass spectra of $\Upsilon(5S)\to \Upsilon(2S)\pi^+\pi^-$
can be well understood \cite{Chen:2011zv}, which shows that two observed $Z_b$ structures play an important role to solve a new puzzle in $\Upsilon(5S)\to \Upsilon(2S)\pi^+\pi^-$. However, we must answer what is the source to generate $Z_b(10610)$ and $Z_b(10650)$. 

Due to the peculiar properties of $Z_b(10610)$ and $Z_b(10650)$, the explanations of the exotic state including molecular state were proposed in Ref. \cite{Sun:2011uh,Bondar:2011ev}. In addition, the cusp effect resulted from the $B\bar{B}^*$ and $B^*\bar{B}^*$ thresholds was given by Bugg in Ref. \cite{Bugg:2011jr}. After these theoretical work, 
we were seeking other mechanisms to explain the $Z_b(10610)$ and $Z_b(10650)$ structures. In Ref. \cite{Chen:2011pv}, Chen and Liu first proposed a new decay mechanism existing in $\Upsilon(5S)$ decay, which is named the initial single pion emission (ISPE) mechanism. With $\Upsilon(5S)\to \Upsilon(mS)\pi^+\pi^-$ as an example, we illustrate the ISPE mechanism, where the emitted pion with continuous energy distribution makes the intermediate $B^{(*)}$ and $\bar{B}^{(*)}$ have low momenta. Hence, $B^{(*)}$ and $\bar{B}^{(*)}$ easily interact with each other to transit into final states
$\Upsilon(mS)\pi$. 

By calculating these $\Upsilon(5S)\to h_b(mP)\pi^+\pi^-$ and $\Upsilon(5S)\to \Upsilon(nS)\pi^+\pi^-$ processes by introducing the ISPE mechanism, we explain why the charged structures near the $B\bar{B}^*$ and $B^*\bar{B}^*$ thresholds can be found in the hidden-bottom dipion decays of $\Upsilon(5S)$. We also indicated that we cannot find the sharp peak close to the $B\bar{B}$ threshold  \cite{Chen:2011pv}.

\section{Charmonium-like structures in dipion decays
}

If the ISPE mechanism is a universal mechanism in heavy quarkonium dipion decays, we naturally extend the ISPE mechanism to study the hidden-charm dipion decays of higher charmonia because of the similarity between charmonium and bottomnium families. We would like to give more predictions for future experiments, which can be as an important test for the ISPE mechanism proposed in Ref. \cite{Chen:2011pv}. 

In Ref. \cite{Chen:2011xk}, the hidden-charm dipion decays of higher charmonia $\psi(4040)$, $\psi(4160)$, $\psi(4415)$ and charmonium-like state $Y(4260)$ were calculated by introducing the ISPE mechanism. Here, we predicted the charged charmonium-like structures near the $D\bar{D}^*$ and $D^*\bar{D}^*$ thresholds in the $J/\psi\pi$, $\psi(2S)\pi$ and $h_c(1P)\pi$ invariant mass spectra. We found that our obtained $h_c(1P)\pi^+$ distribution of $\psi(4160)\to h_c(1P)\pi^+\pi^-$ can depict the CLEO-c measurement of the $h_c(1P)\pi^\pm$
mass distribution from $e^+e^-\to h_c(1P)\pi^+\pi^-$ at $E_{CM}=4170$ MeV \cite{CLEO:2011aa}. Later, Yuan, who represented the Belle Collaboration, gave a talk in "Hadron Structure and Interaction 2011" conference. The comparison between the Belle data and our result of $Y(4260)\to J/\psi\pi^+\pi^-$ were listed, which shows that there is the evidence of enhancement structures in the $J/\psi\pi^\pm$ invariant mass spectrum as predicted by us in Ref. \cite{Chen:2011xk}. 

As predicted in Ref. \cite{Chen:2011xk}, there exists a charged charmonium-like structure near the $D\bar{D}^*$ threshold announced by three experimental analyses. In 2013, the BESIII Collaboration observed a charged charmonium-like structure $Z_c(3900)$ in the $J/\psi\pi^\pm$ invariant mass spectrum of $Y(4260)\to J/\psi\pi^+\pi^-$, which is near the $D\bar{D}^*$ threshold \cite{Ablikim:2013mio}. Belle independently studied the same decay process and claimed the observation of $Z_c(3900)$ \cite{Liu:2013dau}. In Ref. \cite{Xiao:2013iha}, Xiao {\it et al.} analyzed the CLEO-c data of $e^+e^-\to \psi(4160)\to J/\psi\pi^+\pi^-$, and found $Z_c(3900)$ single. 

The observation of $Z_c(3900)$ provides a direct support for the prediction given by the ISPE mechanism. After the observation of $Z_c(3900)$, we carried our further study of $Z_c(3900)$ by combining with the ISPE mechanism in Ref. \cite{Chen:2013coa}, where we considered the interference effects among different mechanisms. Our calculation shows that the $Z_c(3900)$ structure can be well reproduced by analyzing the $J/\psi\pi^+$ and $\pi^+\pi^-$ invariant mass spectra of $Y(4260)\to J/\psi\pi^+\pi^-$. 

\section{More novel phenomena
}

Under the ISPE mechanism, we studied the hidden-charm dipion decays of charmonium-like state $Y(4360)$, where 
$Y(4360)\to J/\psi \pi^+\pi^-$, $\psi(2S) \pi^+\pi^-$, and $h_c(1P)\pi^+\pi^-$ \cite{Chen:2013bha} were considered. 
We also predicted the enhancement structure near $D\bar{D}^*$ and $D^*\bar{D}^*$ thresholds existing in $J/\psi\pi$, $\psi\pi$ and $h_c(1P)\pi$ invariant mass spectra \cite{Chen:2013bha}. 

As an extension of the ISPE mechanism, the initial single chiral particle emission (ISChE) mechanism was proposed in 
Ref. \cite{Chen:2013wca}. Since the pion and kaon can be categorized into chiral particles, we explored the hidden-charm dikaon decays of higher charmonia, which are intriguing processes, where the ISChE mechanism can play an important role in these discussed decays. Via the ISChE mechanism, we have studied the processes $\psi(4415)\to J/\psi K^+K^-$, $Y(4660)\to J/\psi K^+K^-$, and $\psi(4790)\to J/\psi K^+K^-$, where $Y(4660)$ is a charmonium-like state observed in $e^+e^-\to \psi(2S)\pi^+\pi^-$ \cite{Wang:2007ea}, while $\psi(4790)$ is a predicted higher charmonium with quantum number $5S$ in Ref. \cite{vanBeveren:2008rt}. By this study, we predicted charged charmonium-like structures with the hidden-charm and open-strange. Our result indicates that there is no charged charmonium-like structure near the $D_s\bar{D}$ threshold with hidden-charm and hidden-strange in the $J/\psi K^+$ invariant mass distribution of $\psi(4415)\to J/\psi K^+K^-$. However, for $Y(4660)\to J/\psi K^+K^-$, and $\psi(4790)\to J/\psi K^+K^-$ decays, there exist enhancement structures with both hidden-charm and open-strange. These predicted charged charmonium-like structures can be accessible by future experiments, especially BESIII and forthcoming BelleII.  

\section{Summary}

In this talk, we have introduced how we have proposed the ISPE mechanism existing in the hidden-bottom/charm dipion decays of higher bottomonia/charmonia, which is due to the lesson in the study of the hidden-bottom dipion decays of $\Upsilon(5S)$, where the $Z_b$ structures can be understood via the ISPE mechanism. Considering the similarity between the $J/\psi$ and $\Upsilon$ families, the ISPE mechanism was applied to investigate the hidden-charm dipion decays of higher charmonia or charmonium-like states, we also predicted the charged charmonium-like structures near 
the $D\bar{D}^*$ and $D^*\bar{D}^*$ thresholds, and introduced the comparison of these results with the experimental measurements. The $Z_c(3900)$ observation by BESIII and Belle confirmed our prediction, which is an important test of the ISPE mechanism. For further testing the ISPE mechanism, we applied the ISEP mechanism to hidden-charm dipion decays of charmonium-like state $Y(4360)$, where the charged charmonium-like structures near the $D\bar{D}^*$ and $D^*\bar{D}^*$ thresholds were predicted. As an important extension of the ISPE mechanism, we proposed the ISChE mechanism to study the hidden-charm dikaon decays of higher charmonia. Some predictions of charged enhancement structures with both hidden-charm and open-strange were also given. 

Under the ISEP and ISChE mechanisms, we have predicted abundant novel phenomena, which are waiting for test from further experiments, especially from BESIII and BelleII. We believe that this serves a good opportunitiy for experimentalists to study charmonium-like states. We expect more experimental progress on this research field that has full of challenges.

Finally we give a list of processes we have worked on with the ISChE mechanism in Table \ref{Tab:ISChE}.

\begin{table}
\caption{List of processes using the ISChE mechanism. Values in parenthses with a question mark in front are the predicted values.}
\label{Tab:ISChE}
\begin{tabular}{c|c|c|c}
\hline
\rm{Initial~State} & {\rm{Intermediates}}& {\rm{Peak}}& {\rm{Final~State}}\\
\hline
${\psi (4040)}$~\cite{Chen:2011xk}& ${{D^*}\bar D + H.c.}$& {?(3860)}& ${\psi {\rm{(2S)}}{\pi ^ + }{\pi ^ - }/{h_c}(1P){\pi ^ + }{\pi ^ - }}$ \\
\hline
${\psi (4160)}$~\cite{Chen:2011xk}& ${{D^*}\bar D + H.c.}$& {?(3860)}& ${J/\psi {\pi ^ + }{\pi ^ - }/{h_c}(1P){\pi ^ + }{\pi ^ - }}$\\
\hline
${\psi (4415)}$~\cite{Chen:2011xk}& $\begin{array}{l}
{D^*}\bar D + H.c.\\
{D^*}{{\bar D}^*}\\
{D^*}{{\bar D}^*}
\end{array}$& $\begin{array}{l}
?(3860)\\
?(4016)\\
?(4016)
\end{array}$& $\begin{array}{l}
\psi /J{\pi ^ + }{\pi ^ - }\\
\psi (2S){\pi ^ + }{\pi ^ - }\\
{h_c}(1P){\pi ^ + }{\pi ^ - }
\end{array}$\\
\hline
${Y(4260)}$~\cite{Chen:2011xk}& ${{D^*}\bar D + H.c.}$& {?(3860)}& ${J/\psi {\pi ^ + }{\pi ^ - }/{h_c}(1P){\pi ^ + }{\pi ^ - }}$\\
\hline
${\Upsilon (5S)}$~\cite{Chen:2011pv}& $\begin{array}{l}
{B^*}\bar B + H.c.\\
{B^*}{{\bar B}^*}\\
{B^*}\bar B + H.c.\\
{B^*}{{\bar B}^*}
\end{array}$& $\begin{array}{l}
\Upsilon (10610)\\
\Upsilon (10650)\\
\Upsilon (10610)\\
\Upsilon (10650)
\end{array}$& $\begin{array}{l}
\Upsilon (1,2,3S){\pi ^ + }{\pi ^ - }\\
\Upsilon (1,2,3S){\pi ^ + }{\pi ^ - }\\
{h_b}(1,2P){\pi ^ + }{\pi ^ - }\\
{h_b}(1,2P){\pi ^ + }{\pi ^ - }
\end{array}$\\
\hline
${\Upsilon (11020)}$~\cite{Chen:2011pu} & $\begin{array}{l}
{B^*}\bar B + H.c.\\
{B^*}{{\bar B}^*}
\end{array}$& $\begin{array}{l}
?(10605)\\
?(10650)
\end{array}$& $\begin{array}{l}
\Upsilon (1,2S){\pi ^ + }{\pi ^ - }/{h_b}(1,2P){\pi ^ + }{\pi ^ - }\\
\Upsilon (1,2S){\pi ^ + }{\pi ^ - }/{h_b}(1,2P){\pi ^ + }{\pi ^ - }
\end{array}$\\
\hline
${\Upsilon (10860)}$ ~\cite{Chen:2011pu} & ${{B^*}\bar B + H.c.}$& {?(10605)}& ${\Upsilon (1,2S){\pi ^ + }{\pi ^ - }}$\\
\hline
${Y(2175)/\phi (1680)}$~\cite{Chen:2011cj} & ${{K^*}(892)\bar K + H.c.}$& {?(1386)}& ${\phi (1020){\pi ^ + }{\pi ^ - }}$\\
\hline
${\Upsilon(5S)}$ ~\cite{Chen:2011aa}  & $\begin{array}{l}
B{{\bar B}^*}\\
{B^*}{{\bar B}^*}
\end{array}$& $\begin{array}{l}
{{\rm{Z}}_b}(10610)\\
{{\rm{Z}}_b}(10650)
\end{array}$& $\begin{array}{l}
B{{\bar B}^*}\pi \\
{B^*}{{\bar B}^*}\pi
\end{array}$ \\
\hline
${\psi (4415)}$ ~\cite{Chen:2011aa} & $\begin{array}{l}
{D^{*0}}{D^ - }\\
{D^{*0}}{D^{* - }}
\end{array}$& $\begin{array}{l}
?(3877)\\
?(4017)
\end{array}$& $\begin{array}{l}
{D^{*0}}{D^ - }{\pi ^ + }\\
{D^{*0}}{D^{* - }}{\pi ^ + }
\end{array}$\\
\hline
${\psi (4160)}$ ~\cite{Chen:2011aa} & ${{D^{*0}}{D^ - }}$& {?(3877)}& ${{D^{*0}}{D^ - }{\pi ^ + }}$\\
\hline
${Y(4260)}$ ~\cite{Chen:2013coa} & ${{D^{*0}}{D^ - }/{D^{*0}}{D^{* - }}}$& ${{Z_c}(3900)}$& ${J/\psi {\pi ^ + }{\pi ^ - }}$\\
\hline
${\psi (4360)}$ ~\cite{Chen:2013bha} & ${{D^*}\bar D + H.c.}$& {?(3860)}& ${J/\psi {\pi ^ + }{\pi ^ - }/{h_c}(1P){\pi ^ + }{\pi ^ - }}$\\
\hline
${Y(4660)}/\psi(4790)$ ~\cite{Chen:2013wca} &$\begin{array}{l}
{D^*}{\bar D_s}+{D}{\bar D_s^*}+H.c.\\
{D^{*}}{\bar D_s^{*}}+H.c.
\end{array}$& ?(3976/3977) &
$J/\psi K^+K^-$\\
\hline
${Y(4660)}/\psi(4790)$ ~\cite{Chen:2013axa} &$\begin{array}{l}
{D^{(*)}}{\bar D^{(*)}}/{D_s^{(*)}}{\bar D_s^{(*)}}
\end{array}$& ?(4018/4081/4225) &
$J/\psi \eta\eta$\\
\hline
${Y(4260)}$ ~\cite{Wang:2013qwa} & ${{D^{*0}} D^{*-}}$& ${Z_c(4025)}$& ${{D^{*0}} D^{*-}\pi^+}$\\
\hline 
\end{tabular}
\end{table}

\bibliographystyle{aipproc}   

\bibliography{sample}


\IfFileExists{\jobname.bbl}{}
 {\typeout{}
  \typeout{******************************************}
  \typeout{** Please run "bibtex \jobname" to optain}
  \typeout{** the bibliography and then re-run LaTeX}
  \typeout{** twice to fix the references!}
  \typeout{******************************************}
  \typeout{}
}

\end{document}